\documentstyle[11pt,epsf]{article}
\newcommand{\ewxy}[2]{\setlength{\epsfxsize}{#2}\epsfbox[10 60 640 570]{#1}}

\let\De=\Delta

\let\eps=\epsilon

\let\La=\Lambda

\let\si=\sigma

\def\to{\rightarrow}
\let\p=\partial
\let\<=\langle
\let\>=\rangle

\def\eqn#1{(\ref{#1})}  
\def\e{ {\rm e} }

\def\beq{\begin{equation}}
\def\eeq{\end{equation}}
\def\ba{\begin{array}}
\def\bea{\begin{eqnarray}}
\def\ea{\end{array}}
\def\eea{\end{eqnarray}}

\def\comment#1{ \hbox{[{\it Comment suppressed here.}\/]} }
\def\hide#1{}
\def\o{\over}

\def\Ord{ {\rm O} }

\def\k{{\bf k}}
\def\r{{\bf r}}

\def\IR{\relax{\rm I\kern-.18em R}}
\def\IN{\relax{\rm I\kern-.18em N}}
\def\IB{\relax{\rm I\kern-.18em B}}
\def\IE{\relax{\rm I\kern-.18em E}}
\def\ZZ{\relax{\sf Z\kern-.4em Z}}

\def\TT{\mathchoice
       {\sf T\kern-0.52 em{T}}{\sf T\kern-0.52 em{T}}
       {\sf T\kern-0.40 em{T}}{\sf T\kern-0.40 em{T}}}
\def\IP{\mathchoice
       {\sf I\kern-0.14 em{P}}{\sf I\kern-0.14 em{P}}
       {\sf I\kern-0.11 em{P}}{\sf I\kern-0.11 em{P}}}
\def\id{1\kern-.25em {\rm l}}

\def\ontopss#1#2#3#4{\raise#4ex \hbox{#1}\mkern-#3mu {#2}}

\newcommand{\skipover}[1]{}
\newcommand{\nn}{\nonumber \\}

\def\MSb{{\overline{{\rm MS}}}}

\def\s={\! = \!}        

\def\l={\,=\,}
\def\+{\,+\,}
\def\-{\,-\,}

\newdimen\pmboffset
\def\oldpmb#1{\setbox0=\hbox{#1}%
 \copy0\kern-\wd0
 \kern\pmboffset\raise 1.732\pmboffset\copy0\kern-\wd0
 \kern\pmboffset\box0}

\def\appendix{\par                              
    \setcounter{section}{0}                     
    \setcounter{subsection}{0}
    \renewcommand{\theequation}{\Alph{section}.\arabic{equation}}
    \renewcommand{\thesection}{Appendix \Alph{section}
                \setcounter{equation}{0}  } 
}

\catcode`\@=11

\def\section{
\setcounter{equation}{0}        
\@startsection {section}{1}{\z@}{-3.5ex plus -1ex minus 
  -.2ex}{2.3ex plus .2ex}{\Large\bf}}
\renewcommand{\theequation}{\arabic{section}.\arabic{equation}}

\def\subsection{\@startsection{subsection}{2}{\z@}{-3.25ex plus -1ex minus 
 -.2ex}{1.5ex plus .2ex}{\normalsize\bf}}

\def\subsubsection{\@startsection{subsubsection}{3}{\z@}{-3.25ex plus
 -1ex minus -.2ex}{1.5ex plus .2ex}{\normalsize}}

\def\@eqnnum{%
\savebox{\eqnumb}{\rm (\theequation)}%
\settowidth{\numblen}{\usebox{\eqnumb}}%
\makebox[\numblen][l]{\usebox{\eqnumb}~~~\usebox{\eqlabel}}}

\catcode`\@=12

\newsavebox{\eqlabel}

\catcode`\@=11
\newlength{\numblen}
\newsavebox{\eqnumb}
\def\@eqnnum{%
\savebox{\eqnumb}{\rm (\theequation)}%
\settowidth{\numblen}{\usebox{\eqnumb}}%
\makebox[\numblen][l]{\usebox{\eqnumb}~~~\usebox{\eqlabel}}%
}
\catcode`\@=12

\newenvironment{equationwithlabel}[1]{ %
  \savebox{\eqlabel}{#1}
  \begin{equation}\label{#1} }{\end{equation}\savebox{\eqlabel}{~}}

\newcommand{\beql}[1]{\begin{equationwithlabel}{#1}}
\newcommand{\eeql}{\end{equationwithlabel}}

\newenvironment{eqnarraywithlabel}[1]{ %
  \savebox{\eqlabel}{#1}
  \begin{eqnarray}\label{#1} }{\end{eqnarray}\savebox{\eqlabel}{~}}

\newcommand{\beal}[1]{\begin{eqnarraywithlabel}{#1}}
\newcommand{\eeal}{\end{eqnarraywithlabel}}
\begin{document}

\begin{flushright}
FSU-SCRI-97-122 \\
November 1997 \\
\end{flushright}
\vskip 6mm
\begin{center}

{\LARGE \bf Accurate Scale Determinations for\\[2.5mm]
            the Wilson Gauge Action}

\vskip 9mm
{\normalsize R.G. Edwards, U.M. Heller and T.R. Klassen\\[1.2mm]
             SCRI, Florida State University\\[1mm]
             Tallahassee, FL 32306-4130, USA}
\vskip 8mm

{\normalsize \bf Abstract}

\vskip 7mm

\begin{minipage}{5.0in}  
{\small 
Accurate determinations of the physical scale of a lattice action
are required to check scaling and take the continuum limit. 
We present a high statistics study of the static potential for the
SU(3) Wilson gauge action on coarse lattices ($5.54 \leq \beta \leq 6.0$).
Using an improved analysis procedure we determine the string tension
and the Sommer scale $r_0$ (and related quantities) to 1\% 
accuracy, including all systematic errors.
Combining our results with earlier ones on finer lattices,
we present parameterizations of these quantities 
that should be accurate to about 1\% for  $5.6 \leq \beta \leq 6.5$. 
We estimate the $\La$-parameter of quenched QCD
to be $\La_\MSb \l= 247(16)~{\rm MeV}$.
}

\end{minipage}
\end{center}
\vskip 8mm

\renewcommand{\thepage}{\arabic{page}}
\setcounter{page}{1}


\section{Introduction}

One of the technical but very important parts of a lattice simulation
is the determination of the physical scale of the lattice. 
Only with accurate scale determinations can one check if the ratio of
an observable to the physical scale has the theoretically expected
scaling violations (e.g.~order $a^2$ for the Wilson gauge action).
This is required for the accurate and reliable continuum extrapolation
of observables, and for their conversion into physical units to compare
with (or predict) experiment.

Accurate scale determinations have been given new impetus by the
recent progress in the Symanzik improvement program 
(see e.g.~\cite{LAT97,LAT96}). In particular, the rather
non-trivial, non-perturbative determination~\cite{ALPHA} 
(cf.~also~\cite{EHKlat97,EHK})
of the $\Ord(a)$ coefficient in the Sheikholeslami-Wohlert (SW) quark 
action coupled to the Wilson gauge action has to be verified by
accurate scaling checks. Only then would we feel confident to proceed 
with other steps in the improvement program.

It is known that the rho mass is a sensitive indicator of scaling
violations, in particular of the $\Ord(a)$ violations of chiral
symmetry that are generically present in a Wilson-type quark action
like the SW action. The rho mass is an ideal quantity to use in scaling 
checks. Therefore we would rather not use it  to set the scale,
as is sometimes done. Instead we have to determine some other quantity
sufficiently accurately to set the scale. 

In pure gauge theory (or the quenched approximation of full QCD) the
string tension $\si$ obtained from the static potential is such a quantity 
--- in principle. In practice technical problems have often interfered
with accurate determinations of this quantity. This is especially
true on coarse lattices, $\beta < 6.0$ for the Wilson gauge 
action,\footnote{We use the standard
notation $\beta\equiv 6/g^2$ to parameterize the bare gauge coupling
of the SU(3) Wilson plaquette action.}
where, otherwise, scaling checks could be performed most accurately.

To our surprise we found that no accurate string tensions are known
for the Wilson gauge action for $\beta < 6.0$. The newest results
we are aware of    were presented in refs.~\cite{MTc,SB}.
The errors are relatively large, and combined with string tensions on
finer lattices the results do not really seem to lie on a smooth curve.

We have therefore decided to perform high statistics simulations of 
the static potential for the Wilson gauge action with $\beta \leq 6.0$,
to determine the string tension and Sommer-type scales~\cite{Sommer}
(the latter are also well-defined for full QCD).
In sect.~2 we describe procedures to accurately calculate these quantities 
from the static potential. Sect.~3 gives the details of our
simulations and the numerical results. 
In sect.~4 we present an accurate parameterization of
our results for all $\beta$ in the range $5.6\leq \beta \leq 6.5$.
In sect.~5 we discuss the issue of scaling and extracting the $\La$-parameter
of QCD from the measured string tensions.
We conclude in sect.~6.

\section{Scale Determination with the Static Potential}

A standard method of fitting lattice Monte Carlo
data for the static potential between an infinitely heavy quark
and anti-quark  is to use an 
ansatz of the form~\cite{Michael}
\beq\label{Vansatz}
 V(\r) \l= V_0 \+ \si r \- e \, \left[{1\o \r}\right] \+ 
               l \left (\, \left[{1\o \r}\right] \- {1\o r}\, \right) \, .
\eeq
Here $[{1\o \r}]$ denotes the tree-level lattice Coulomb term,
viz,
\beq
  \left[{1\o \r}\right] \l= 4\pi \int {d^3\k\o (2\pi)^3} ~
         \cos(\k\!\cdot \! \r) ~ D_{00}(0,\k) \, ,
\eeq
where $D_{00}(k)$ is the time-time component of the gluon propagator
for the gauge action in question.
Note that $l$ is an effective parameter introduced to model lattice 
artifacts beyond tree-level.

Clearly, eq.~\eqn{Vansatz} is not a fundamental ansatz, since
neither the running of the QCD coupling nor the lattice artifacts
are taken into account properly (for a more ambitious attempt
see~\cite{TKpotl}). This is a problem if one tries to extract
$\La_{{\rm QCD}}$ or other (intermediate- to) short-distance
aspects of the running coupling from potential data
via an ansatz of the above or similar form.

We are here not interested in short-distance properties of the
potential. Instead, we would like to extract a physical scale
from the (intermediate- to) long-distance part of the potential.

Let us first consider determining the string tension $\si$, which,
according to the flux tube picture and MC studies, is a well-defined
quantity in pure gauge theory. 
When trying to determine only the string tension, the problems
of the ansatz~\eqn{Vansatz} at short distances become largely
irrelevant. In practice we will always leave out the first few
lattice points and check if the string tension stabilizes as we 
include in the fit only points with $r \geq r_{{\rm min}}$,
with increasing $r_{{\rm min}}$.

According to the string picture the potential in pure gauge theory
behaves for large distances as
\beq
  V(r) \l= V_0 \+ \si r \- {e_{{\rm IR}} \o r} \+ \ldots .
\eeq
If the transverse fluctuations of the QCD flux tube are described
by the simplest bosonic string theory, then the ``IR charge'' takes
on the universal value $e_{{\rm IR}} = \pi/12$ in four space-time
dimensions~\cite{eIR}.
Like others, we have observed that even with very good statistics (and 
smearing, etc) the intermediate- and long-distance part of lattice potential
data can indeed always be described by an ansatz of the form~\eqn{Vansatz}
with $e=\pi/12$.

\vskip 1mm

To determine the string tensions we therefore perform three types of
fits:
\begin{itemize}
\item 4-parameter fits to~\eqn{Vansatz}.

\item 3-parameter fits to~\eqn{Vansatz} with $e=\pi/12$ fixed.

\item 2-parameter fits to~\eqn{Vansatz} with $e=\pi/12$  and $l=0$ fixed.

\end{itemize}
As $r_{{\rm min}}$ increases for the various fits, they should reach
consistent values for the string tension. This asymptotic value is expected
to emerge first for the 4-parameter fit, next for the 3- and finally
for the 2-parameter fit.  The demand that all three fits give a consistent
value should allow an accurate and reliable determination of the string 
tension. We should remark that similar ideas were used in the
string tension determinations by the Bielefeld group~\cite{BIsigma}.

Although the method just sketched works quite
well, as we will see, there are clearly drawbacks to using the string
tension to set the scale. First of all, of course, in full QCD 
the string tension is not well-defined, due to string breaking. Secondly,
due to its long-distance nature, one needs the 
potential in a region where errors are becoming large.

To avoid these problems Sommer~\cite{Sommer} suggested to define an 
intermediate distance scale $r_c$ via the {\it force} between a heavy 
quark and anti-quark,
\beq
        r_c^2 \, V'(r_c) \l= r_c^2 \, F(r_c) \, \stackrel{!}{=} \, c
\eeq
for some suitably chosen real number $c$.
The question is how to implement this idea in practice to extract $r_c$
from lattice data.

One approach is to define the force and its error via discrete
(correlated) differences of pairs of points,
\beq
  F(r_I) ~\equiv~ {| V(\r_1) - V(\r_2)| \o | \r_1 - \r_2 | } \, ,
\eeq
where $r_I$ is chosen~\cite{Sommer} to eliminate tree-level
lattice artifacts,
\beq
  {1\o r_I^2} ~\equiv~  -{| [{1\o \r_1}] - [{1\o \r_2}] | \o
                          |       \r_1   -       \r_2   | }  \, .
\eeq
In practice one finds that the force values from different pairs of points 
around $r_c$ still do not fall on a smooth curve, due to 
remaining lattice artifacts. When interpolating the force to obtain $r_c$
one therefore ends up with relatively large errors.

One can do better (see e.g.~\cite{SESAM})
 by defining $r_I$ via the {\it corrected}
(more continuum-like) potential data,
\beq
  V_c(\r) ~\equiv~ V_{{\rm MC}}(\r) \+ 
        (e-l)\, \left(\left[{1\o\r}\right]-{1\o r}\right) \, ,
\eeq
as
\beq\label{rSESAM}
  {\p V_c\o \p r}(r_I) \l= \si + {e\o r_I^2}
  ~ \stackrel{!}{=} ~ {| V_c(\r_1) - V_c(\r_2)| \o | \r_1 - \r_2 | } \, ,
\eeq
where $\si$, $e$ and $l$ are taken from a fit to the ansatz~\eqn{Vansatz}.
By (slight) abuse of notation we are also using  
$V_c(r) \equiv V_0 + \si r -e/r$ on the left side of 
eq.~\eqn{rSESAM} to denote the continuum part of~\eqn{Vansatz}.

Taking the last idea to its logical conclusion we will use the 
following method to determine $r_c$: Perform a correlated fit of 
the potential data to an ansatz of the form~\eqn{Vansatz} --- but only 
{\it locally}, using data between some $r_{{\rm min}}$ and $r_{{\rm max}}$
close to $r_c$. We then obtain $r_c$ from the continuum part of the ansatz
as
\beq
         r_c ~=~ \sqrt{ { c - e\o \si } } ~ .
\eeq
Since we are not performing a global fit of the potential data the
precise from of the (continuum part of the) ansatz should not be 
crucial.\footnote{One would therefore expect the general
4-parameter fit to be somewhat over-parameterized. Indeed, we find
that for $r_c$ fits  $e$ and $\si$ are strongly (anti-) correlated.}
We can estimate the systematic error of $r_c$ by varying $r_{{\rm min}}$
and $r_{{\rm max}}$, and by fixing $e=\pi/12$.
Similarly, we can get an idea of the effect of lattice
artifacts beyond tree-level by comparing free $l$ with $l=0$ fits.

The nice feature of this strategy is that we can avoid taking the
numerical derivatives required for the force, while still using only
local information about the potential and being able to take lattice
artifacts beyond tree-level into account. The fits required are of the
same kind that one performs to determine the string tension; one
just has to consider smaller intervals $[r_{{\rm min}}, r_{{\rm max}}]$
around $r_c$.

\section{Simulations and Results}

We performed simulations at five couplings $\beta \leq 6.0$ for
the SU(3) Wilson plaquette action. Our updating algorithm alternates
microcanonial over-relaxation and heatbath steps acting on SU(2)
subgroups~\cite{CabMar}
(typically in a 3:1 ratio per sweep). In all cases we generated several 
thousand (essentially) independent configurations on lattices whose
spatial extent is at least 1.5~fm; cf.~table~\ref{tab:simpar}.
We should remark that a subset of the $32\!\cdot \! 16^3$ configurations
at $\beta\s=5.7$ and $5.85$ are also being used for spectrum 
calculations with the non-perturbatively improved SW quark 
action~\cite{EHK}.

\begin{table}[tb] \centering
\begin{tabular}{ | l | c | c | c | }
\hline
~$\beta$~~  & ~Volume~ & $L$~(fm) & ~$N_{{\rm confgs}}$~ \\ \hline  
6.0   & $16^3\! \cdot\!32$ & 1.5 & 4000      \\               
5.85  & $16^3\! \cdot\!32$ & 2.0 & 8000      \\               
5.7   & $16^3\! \cdot\!32$ & 2.6 & 4000      \\               
5.6   &  $12^4$            & 2.6 & 2000      \\               
      &  $8^4$             & 1.7 & 4630      \\               
5.54  &  $12^4$            & 2.9 & 2000      \\               
\hline
\end{tabular}      
\vskip 2mm
\caption{Volume, spatial extent
         and number of configurations used in our simulations.}
\label{tab:simpar}
\vskip 5mm
\end{table}

The static potential is obtained in the standard manner from (smeared)
Wilson loops $W(\r,t)$ via the effective local masses
\beq\label{Vt}
     V_t(\r) \,\, \equiv \,\, \ln \, { W(\r,t) \o W(\r,t+1) } \, .
\eeq
To reach plateaus for small $t$ we perform APE smearing~\cite{APE}
on the spatial link variables. This amounts to adding the spatial
staples to a suitable multiple of the given link,
\beq\label{APE}
   U_k(x) ~\to~ \alpha \, U_k(x) \+ 
\sum_{l:l\neq k} U_l(x) \, U_k(x+\hat{l}) \, U_{-l}(x+\hat{l}+\hat{k}) \, ,
\eeq
and projecting back onto SU(3). This procedure is then iterated
$n$ times.

Note that (modulo details of the SU(3) projection) the above procedure
is essentially equivalent to a more intuitive
``covariant gaussian smearing'' in terms of the
transverse laplacian $\De_\perp$,
\beq\label{covgauss}
  U_k(x) ~\to~ (1+\eps \De_\perp)^n ~ U_k(x) 
       ~\approx~ \e^{\eps \, n \, \De_\perp} ~ U_k(x) \, .
\eeq
In momentum space the exponential becomes a gaussian, at least in the
free case. The relation between $\eps$ and $\alpha$ is\footnote{We do 
not write an exact equality here,
because of   possibly different implementations of the projection,
and the question of whether one uses the initial or the 
``updated'' gauge field in the covariant laplacian.}
$\eps \approx 1/(4+\alpha)$.

This way of writing the smearing
is useful, because it shows that as long as $\eps$ is small
the actual amount of smearing will only depend on the product $\eps \, n$.
Dimensional analysis then also shows that the same amount of smearing, 
in physical units, corresponds to scaling $\eps \, n \propto 1/a^2\,$ as
one varies the lattice spacing. One therefore has to optimize the smearing
only at one coupling and can then use this scaling to find a (near-)
optimal smearing at other lattice spacings.\footnote{The situation is
       more complicated for improved gauge actions, due to
the contributions from ``ghost states'' to correlation functions at
small times.
}
We find that $\eps \, n \approx 4$ is a good smearing for $\beta=6.0$.
Even for large $r$ ground state overlaps are then at least 85\%.

It is important to have off-axis $\r$ to accurately determine
the string tension and $r_c$. 
We measured the potential at multiples of the
lattice vectors $(1,0,0)$, $(1,1,0)$, $(1,1,1)$ and $(2,1,0)$.\footnote{For
the simulations on $8^4$ lattices at $\beta=5.6$ 
we measured several more $\r$ values. This case is
also exceptional in that we used an updating algorithm that alternates
over-relaxation and Metropolis steps.}

Our fitting procedure was sketched in sect.~2. For a given fit, i.e.~for
a given ansatz, $t$ in eq.~\eqn{Vt}, as well as
$r_{{\rm min}}$ and $r_{{\rm max}}$,
we use a single elimination jackknife procedure (after suitable binning of
the data) to obtain the errors of all quantities of interest.
The final value and error we quote is obtained as follows.
For each fit ansatz (with 2, 3, or 4 parameters, cf.~sect.~2) and $t$
we consider the best few fits (five or less, say), 
that satisfy a minimal ``goodness''
criterion. Our ``goodness'' function is typically the product of the
confidence level and the number of degrees of freedom of the fit.
We then combine all the ``good'' fits with a weight that depends on
the goodness and the error of the quantity in question.

This procedure is semi-automated to eliminate subjective bias.
It is hard to completely automate this process. 
For example, it can happen (e.g.~if the smearing is not optimal) that
fits from neighboring pairs of time slices seem roughly
equally ``good'', but differ significantly in their estimate of the
quantity under consideration.  In such cases it seems preferable 
to invoke human intelligence and experience to decide what weight
to assign to the different fits.

Overall our analysis procedure is conservative, because we
do not simply take the one ``best'' fit according to some 
necessarily somewhat subjective  criterion.
By always taking several fits into account we find that our final
numbers and error bars are quite stable under changes of the goodness
function or of the cuts on the minimal required goodness of a fit. 
Our final error
is never smaller than that of any single good fit. In case one fit
is much ``better'' than the others, our error and central value will be 
close to the values of this fit; in general our error bar is significantly
larger, and our central value somewhere between that of the different 
fits.

Using this procedure we have measured the string tension and several
$r_c$. The ``standard'' choice of $r_c$~\cite{Sommer} corresponds to
$c\s=1.65$. Slightly inconsistently, but in accord with accepted
conventions, we will denote this choice by $r_0$. 
It corresponds to $r_0 \approx 0.50$~fm (perhaps a bit less); which is 
around the region where phenomenological potentials are best determined.
On fine lattices it also corresponds to the region where the potential
can be very accurately measured. On lattices coarser than $0.1$~fm
this choice becomes increasingly impractical, however, because 
$r_0$ is then quite small in lattice units. We have therefore decided 
to measure $r_4$ and $r_6$ in addition to $r_0$. We note that
$r_6 \approx 1.0$~fm.

A final remark about the $r_c$ determinations. Recall that in this
case we vary $r_{{\rm min}}$ and $r_{{\max}}$ around $r_c$.
We find, perhaps not too surprisingly, that the dependence on $r_{{\rm max}}$
is always small, in most cases negligible.

\begin{figure}[tbp]
\vskip -12mm
\mbox{ \hspace{-3em}\ewxy{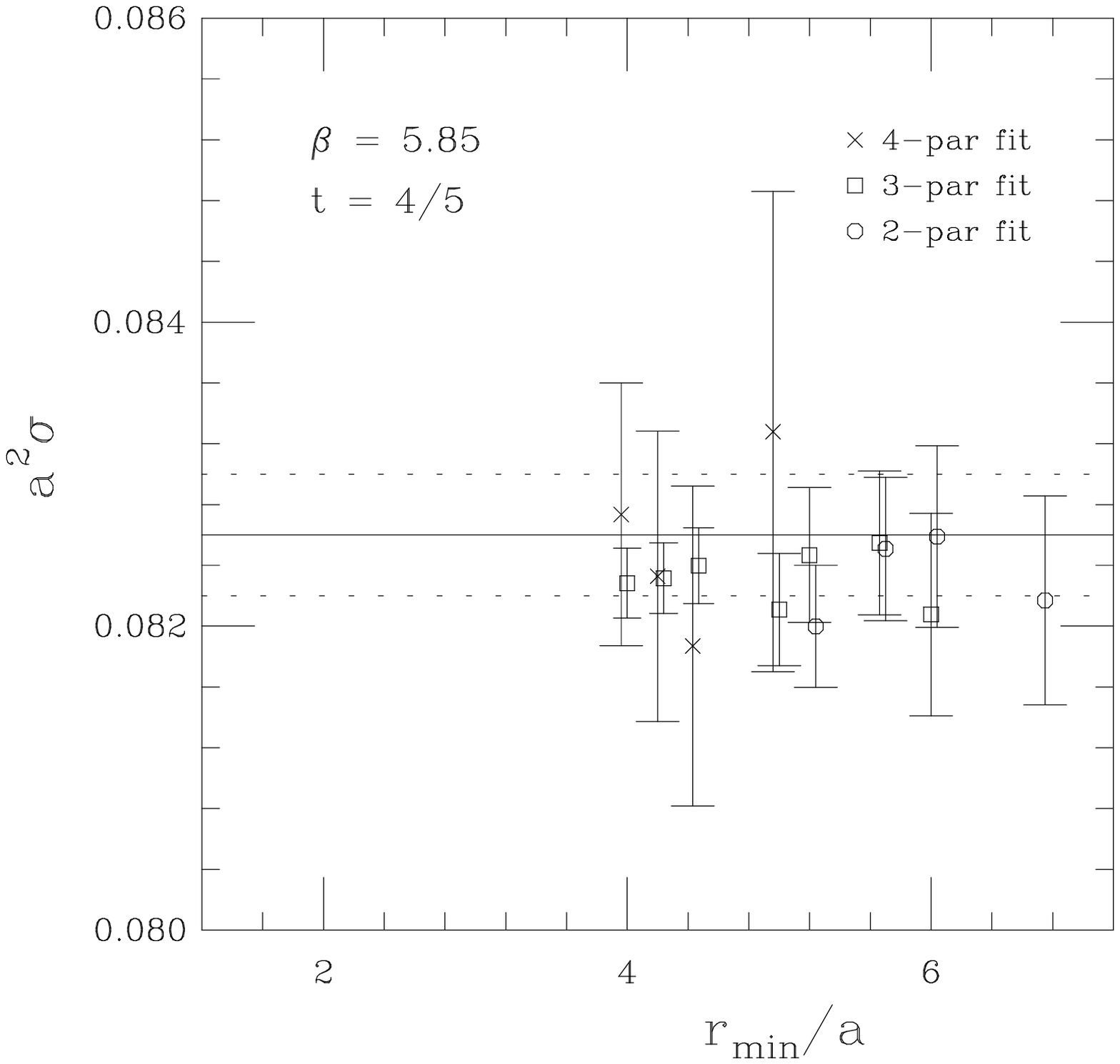}{100mm} 
       \hspace{-6em}\ewxy{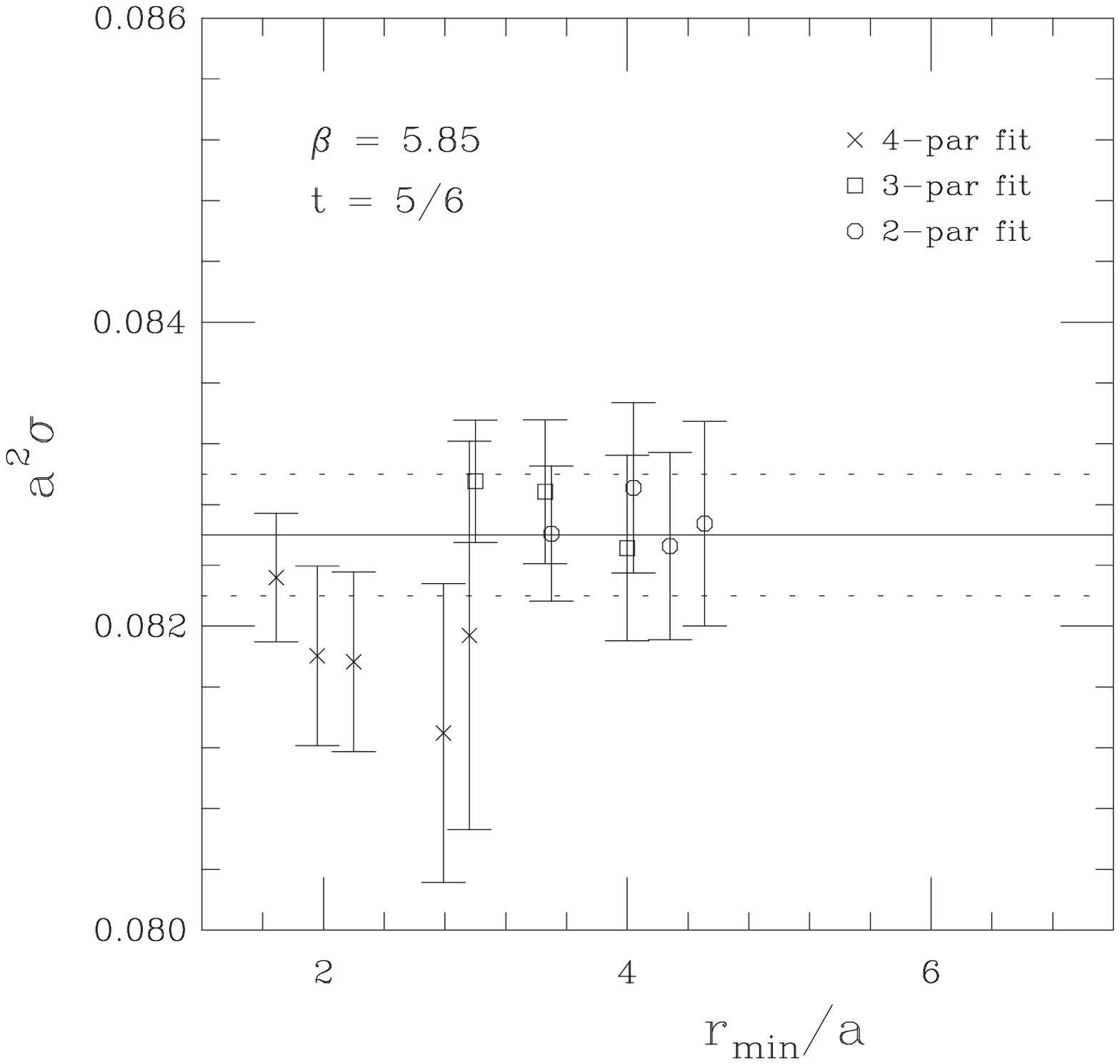}{100mm} } 
\vskip 0mm
\caption{
String tension for $\beta=5.85$ as a function of $r_{{\rm min}}$ for 
different ``good'' fits and using time slices $t=4/5$ and $t=5/6$. The 
solid and dashed lines denote our final value and its error, respectively.
}
\label{fig:sigma585}
\vskip 5mm
\end{figure}

\begin{figure}[tbp]
\vskip -12mm
\centerline{\ewxy{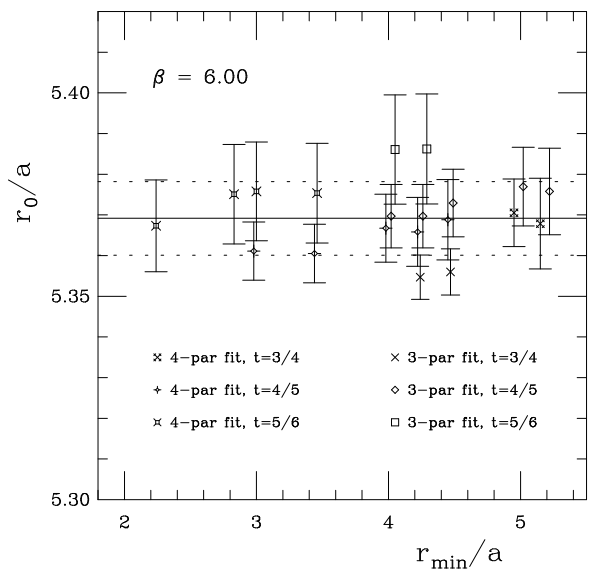}{110mm} }
\vskip 0mm
\caption{
Sommer scale $r_0$ for $\beta=6.0$ as a function of $r_{{\rm min}}$ for 
various ``good'' fits. The solid and dashed
lines denote our final value and its error, respectively.
}
\label{fig:r0_600}
\vskip 4mm
\end{figure}

\begin{figure}[tbp]
\vskip -12mm
\centerline{\ewxy{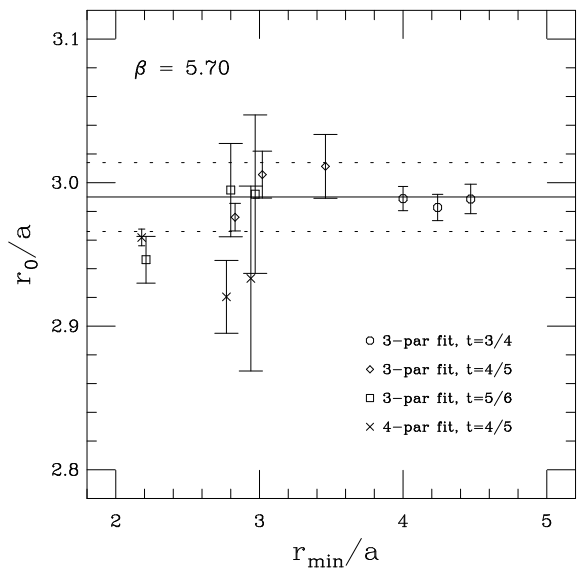}{110mm} }
\vskip 0mm
\caption{
Sommer scale $r_0$ for $\beta=5.70$ as a function of $r_{{\rm min}}$ for 
various ``good'' fits. The solid and dashed
lines denote our final value and its error, respectively. Note that
the 3-parameter fits with $t=3/4$ have $r_{{\rm min}} > r_0$. These
fits were not included in our final error estimate; they are shown
only for illustration.
}
\label{fig:r0_570}
\vskip 4mm
\end{figure}

Some of our results are illustrated in figures~\ref{fig:sigma585}
--~\ref{fig:r0_570}. Note in particular the stability of our determination
of $r_0$ in figure~\ref{fig:r0_570}, 
even for values of $r_{{\rm min}}$ larger than $r_0$ itself!
Such fits are not included in our final analysis, but we 
regard this as a sign of the stable and conservative nature of our
analysis strategy.  
Our final results are collected in table~\ref{tab:oursig}.
The relatively large errors in the $\beta=5.7$ case are presumably 
due to non-optimal smearing; 
we did not follow the guideline described earlier (yet).

\begin{table}[tb] \centering
\begin{tabular}{ | l | l | l | l | l | }
\hline
~$\beta$  & ~~$a\sqrt{\si}$ & ~~$r_0/a$ &  ~~$r_4/a$ & ~~$r_6/a$\\ \hline
5.54  & 0.5727(52) & 2.054(13) & 3.381(21) & 4.187(27)\\
5.6   & 0.5064(28) & 2.344(8)  & 3.814(15) & 4.718(23)\\    
5.7   & 0.3879(39) & 2.990(24) & 4.926(35) & 6.128(42)\\
5.85  & 0.2874(7)  & 4.103(12) & 6.733(17) & 8.340(24)\\
6.0   & 0.2189(9)  & 5.369(9)  & 8.831(21) & 10.89(3)\\
\hline
\end{tabular}      
\vskip 2mm
\caption{Our results for the string tension and Sommer-type
         scales.
}
\label{tab:oursig}
\vskip 5mm
\end{table}

\begin{table}[thb] \centering
\begin{tabular}{ | l | l | l | l | }
\hline
Scale          & $8^4$ (1.7~fm) & $12^4$ (2.6~fm) & ~~Change\\ \hline
$a\sqrt{\si}$  & 0.5004(23)     & 0.5064(28)      & $+1.2\% \pm 0.7\%$\\
$r_0/a$        & 2.355(7)       & 2.344(8)        & $-0.5\% \pm 0.5\%$\\
$r_4/a$        & 3.867(10)      & 3.814(15)       & $-1.4\% \pm 0.5\%$\\
$r_6/a$        & 4.787(14)      & 4.718(23)       & $-1.5\% \pm 0.6\%$\\
\hline
\end{tabular}      
\vskip 2mm
\caption{Check of finite volume effects for $\beta\s= 5.6$.}
\label{tab:finitevolume}
\vskip 5mm
\end{table}

As explained above, we believe that the values quoted include all systematic
errors  --- except possibly for one source we have not discussed so far:
finite volume effects. Previous studies for $\beta\geq 6.0$ indicate that a
spatial extent of about 1.5~fm is sufficient to suppress
finite volume errors in the string tension to the level of about one percent
or less~\cite{BS}. To explicitly check this for our determinations, we 
have repeated the simulations at $\beta\s= 5.6$ on $8^4$ in addition to 
$12^4$ lattices. The comparison is shown in table~\ref{tab:finitevolume}.
It would seem that within our errors there are significant finite volume 
effects on the $1-1.5$\% level, except for $r_0$, which seems to be much less 
affected (not too surprisingly). The sign of the change of the string
tension  is as expected (from earlier studies)
for finite volume effects. That the $r_c$ move in the opposite direction
is also expected.


In any case, for coarse lattices there is no problem in using a 
sufficiently large volume to suppress finite volume effects to 
significantly less than one percent. This is true for the values in 
table~\ref{tab:oursig}, except perhaps for $\beta\s= 6.0$. We will
return to this point in the next section.

\section{Parameterizing the Scale}

The string tension is one of the quantities that has traditionally
been used to check perturbative scaling. It is known that, on a 
quantitative level,  the string tension
follows neither two- or three-loop scaling, at least for
the lattice spacings we are interested in  (we will see this explicitly
below).
Allton~\cite{Allton} 
suggested that this is due to $a^2, g^2 a^2, \ldots, a^4, \ldots$
lattice artifacts, that are unavoidable if one determines the
string tension with the Wilson gauge action.

He therefore proposed~\cite{Allton} to fit the string tension
 to an ansatz of the form
\beq\label{Allton}
(a\sqrt{\si})(g) \l=  
                         f(g^2) ~ ( \,
  1 \+ c_2           \, \hat{a}(g)^2
    \+ c_4           \, \hat{a}(g)^4 \+ \ldots \, 
                            )/{\La\o \sqrt{\si}} \, , \qquad
     \hat{a}(g) \equiv { f(g^2)\o f(g^2(\beta\s=6))} \, , ~
\eeq
in terms of the fit parameters $\La/\sqrt{\si}$, $c_2$, $c_4, \ldots$.
Note that $\hat{a}(g)$ has been normalized to 1 for $\beta\s=6.0$;
we will also use this convention in other coupling schemes considered
later. 
In the above $f(g^2)$ is the universal two-loop scaling function 
of SU(3) gauge theory, 
\beq\label{f2loop} 
 f(g^2) \, \equiv \,
(b_0 g^2)^{-{b_1\o 2b_0^2}} \, \exp(-{1\o 2b_0 g^2}) \, , \qquad
  b_0 = {11\o (4\pi)^2 } \, , \quad  b_1 = {102\o (4\pi)^4} \, .
\eeq
Recall that if scaling holds with a scaling function $f$ in a
scheme $S$, then $\La_S = q f(g^2_S(q))$ is a constant, the 
$\La$-parameter of the scheme.

On theoretical grounds
one might want to add $g^2 \hat{a}(g)^2, \ldots$ corrections to the
terms shown in eq.~\eqn{Allton}, but in practice this does not turn out
to be useful (however, we have checked that the $\La$-parameter obtained
from fits is stable when using such terms in~\eqn{Allton}).

In table~\ref{tab:allsig} we have collected some results for the string
tension and $r_0$. We are not showing data for $\beta\s= 6.6$ and $6.8$,
since they might be affected by finite volume errors~\cite{Gunnar}.
Looking at the different volumes employed for the data shown in 
table~\ref{tab:allsig}, it seems
quite clear that finite volume errors in the string tension are
on the one percent or less level for all $\beta$ (cf.~also our remarks
in sect.~3 and the discussion in the references quoted in the table).

\begin{table}[tb] \centering
\begin{tabular}{ | l | l | l | l | l | }
\hline
~$\beta$  & ~~$a\sqrt{\si}$ &  ~~$r_0/a$ & Volume & Reference  \\ \hline
5.54  & 0.5727(52)  & 2.054(13) & $12^4$     & This work \\
5.6   & 0.5064(28)  & 2.344(8)  & $12^4$     & This work \\
5.7   & 0.3879(39)  & 2.990(24) & $16^3\! \cdot\!32$ & This work \\
5.85  & 0.2874(7)   & 4.103(12) & $16^3\! \cdot\!32$ & This work \\
6.0   & 0.2189(9)   & 5.369(9)  & $16^3\! \cdot\!32$ & This work \\
      & 0.2182(16)  &           & $16^4$ & Perantonis/Michael~\cite{MichPer} \\
      & 0.2184(19)  & $5.35(+2)(-3)$ & $16^4$ & SESAM~\cite{SESAM} \\
      & 0.2209(23)  &           & $32^4$ & Bali/Schilling/Hoeber~\cite{Hoeber} \\
      & 0.2154(50)  & 5.47(11)  & $16^3\!\cdot\! 48$ & UKQCD~\cite{UKQCD1}  \\
6.2   & 0.1610(9)   &           & $32^4$ & Bali/Schilling/Hoeber~\cite{Hoeber} \\
      & 0.1604(11)  & 7.37(3)   & $32^4$ & SESAM~\cite{SESAM} \\
      & 0.1608(23)  & 7.29(17)  & $24^3\!\cdot\! 48$ & UKQCD~\cite{UKQCD1} \\
%
6.4   & 0.1214(12)  & 9.89(16)  & $32^4$ & Bali/Schilling~\cite{BS,Gunnar} \\
      & 0.1218(28)  & 9.75(17)  & $32^3\!\cdot\!64$ & UKQCD~\cite{UKQCD1} \\
6.5   & 0.1068(9)  & 11.23(21)  & $36^4$ & UKQCD~\cite{UKQCD3,BSquote} \\ 
\hline
\end{tabular}      
\vskip 2mm
\caption{Selected world results for the string tension and $r_0$.
}
\label{tab:allsig}
\vskip 5mm
\end{table}

These data are well fitted by an ansatz
of the form~\eqn{Allton} using the two-loop scaling function for the
bare coupling.\footnote{In the next section we will discuss fits in terms
of other couplings, that we consider to be more physical. Here our aim
is the purely pragmatic one of providing an explicit, analytic
parameterization of the data, for which, it turns out, the bare
coupling is just as suited as any other one.}
We obtain for the 12 degree of freedom fit of the string
tension with all data in the table, 
\bea
  {\La\o \sqrt{\si}} & = & 0.01338(12) \, , \nn
                 c_2 & = & 0.195(16) \, , \nn
                 c_4 & = & 0.0562(45) \, ,
\eea
with a confidence level of $Q\s= 0.19$. Adding an $\hat{a}(g)^6$ term
in the ansatz gives a $Q$ of 0.24. For the three-parameter fit of $r_0$
we obtain $Q=0.12$.

When leaving out the $\beta\s=5.54$ point from the fits, the confidence
levels jump up to $Q\s=0.9$ for three-parameter fits of both $\sqrt{\si}$ 
and $r_0$.   This might indicate that the determination of these quantities
at $\beta\s=5.54$ is increasingly tainted by lattice artifacts that
are not modeled with sufficient accuracy by our ansatz.  We find
some signs of this in our potential fits.

\begin{figure}[tbp]
\vskip -12mm
\centerline{\ewxy{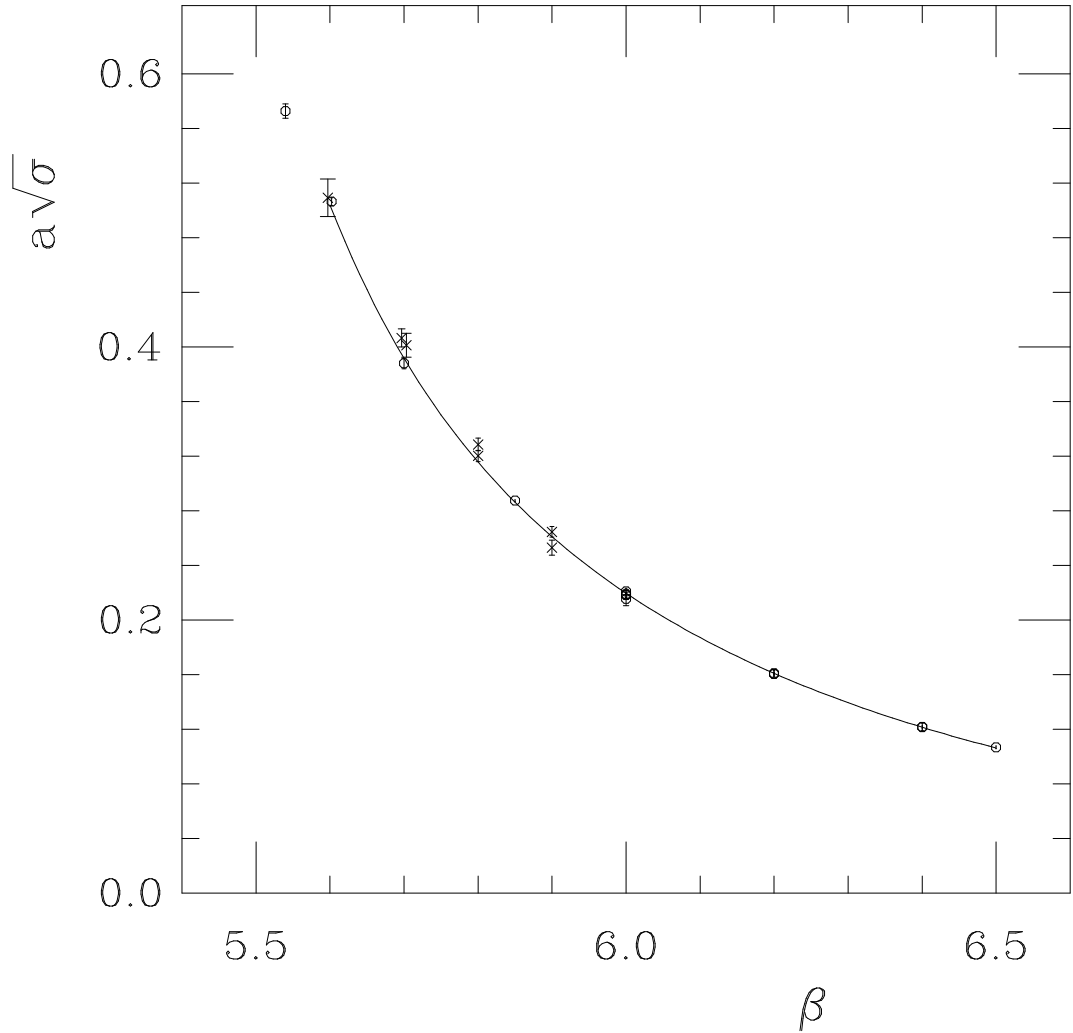}{200mm} }
\vskip 0mm
\caption{
Plot of the parameterization~\protect\eqn{sigparam} of the string
tension as a function of $\beta=6/g^2$. Also shown are the data from
table~\ref{tab:allsig} ($\circ$) that lead to the parameterization, 
as well as earlier results ($\times$) for coarse lattices from 
refs.~\protect\cite{MTc,SB}.
For clarity a few  
data points
have been slightly displaced around their actual $\beta$ value.
}
\label{fig:sig}
\vskip 4mm
\end{figure}

We would like to present
a parameterization of the data that reproduces the central values
within one percent and one standard deviation, whichever is smaller
(considering the tiny error bars in $\sqrt{\si}$ at $\beta\s=5.85$ we 
allow a difference of, say, 1.5 standard deviations there).
This is difficult to achieve with the $\beta\s= 5.54$ point, so we will
do so only for the interval $5.6 \leq \beta \leq 6.5$. 
The following parameterizations satisfy these constraints:
\bea\label{sigparam}
(a\sqrt{\si})(g) & \!=\! & \, f(g^2) ~ ( \,
  1 \+ 0.2731      \, \hat{a}(g)^2
    \- 0.01545     \, \hat{a}(g)^4
    \+ 0.01975     \, \hat{a}(g)^6 \, )/0.01364 \, , \nn
(a/r_0)(g)  & \!=\! & \, f(g^2) ~ ( \,
  1 \+ 0.2106      \, \hat{a}(g)^2
    \+ 0.05492     \, \hat{a}(g)^4 \, )/0.01596 \, .
\eea
The representation of the string tension is shown in 
figure~\ref{fig:sig},
together with the data from table~\ref{tab:allsig} and results obtained
earlier for $\beta < 6.0$.

\section{Scaling and $\La_\MSb$}

One can consider taking the above much more seriously than
just as a way of parameterizing the data. That is, one can try
to extract the $\La$-parameter of QCD from it.  
The statistical uncertainties
of the fit parameters are very small, about 1\% for $\La/\sqrt{\si}$
and slightly larger for $\La r_0$.
However, before taking this seriously as a determination of 
the $\La$ parameter, we have to consider systematic errors.

The most important question is whether the $\La$-parameter extracted
from a two-loop fit with the bare coupling is stable when going to
three-loops, and when considering other schemes for the coupling.
It is believed that perturbative scaling holds better
when using an ``improved'' coupling~\cite{LM}. The precision data
for the string tension that are now available for a wide range
of lattice spacings provide an excellent testing ground of these ideas.

We will consider the
$E$-scheme~\cite{Escheme}, the $V$-scheme~\cite{Vscheme,LM}, and 
the $\MSb$ scheme.
In the $E$-scheme ($E$ for energy) one uses the leading perturbative 
expansion of the expectation value of the plaquette, $W_{11}$, 
to define $g_E^2$ for SU($N$) gauge theory via
\beq\label{W11expn}
  1 \- W_{11} \l= w_1 \, g^2 \+ w_2 \, g^4 \+ \ldots
    \, \equiv \, w_1 g^2_E \, ,  \qquad w_1 = {N^2 -1 \o 8 N } \, .
\eeq
In practice one uses the value of the plaquette measured in a Monte
Carlo simulation on the left hand side.

To obtain the coupling in the $V$-scheme, originally defined via the
static potential, we use the expansion of the logarithm of the plaquette
to second order to first define the ``plaquette coupling'' 
$g^2_P(q)$~\cite{LM}. Using the measured value of the plaquette,
$g^2_P(q)$ is obtained by solving, for pure SU($N$),
\beq\label{logW}
 -\ln W_{11} ~=~ w_1 \, g^2_P(q) \, \left[ 1 \- 
      2b_0 \ln\left({6.711706\o a q}\right) \, g^2_P(q) \right] \, .
\eeq
The optimal scale $q$ to extract the coupling, according to~\cite{LM},
is given by the value $q^\star = 3.4018/a\,\,$\cite{LM,TKpotl}.

By construction $g^2_P = g^2_V + \Ord(g_V^6)$. The exact coefficient of
the third order term is now also known (cf.~\cite{W3loop,LW3loop,Alles} 
with the last missing ingredient recently provided in~\cite{V3loop}), so 
that we can extract $g^2_V$ to third order from $g^2_P$ defined 
above:\footnote{Since the third order expansion of the plaquette is now
also known in terms of $g^2_V$, one could try to obtain the latter directly
from the plaquette, without the detour via $g^2_P$. However, it turns
out that the cubic equation for $g^2_V$ has no physical, i.e.~real and
positive, solutions for $\beta\leq 5.7$. The detour via $g^2_P$ is
therefore to be preferred.}
\beq\label{gVofP}
      g^2_V(q^\star) \l= g^2_P(q^\star)
 \, \left [ 1 \+ {2.814089\o (4\pi)^2} \, g^2_P(q^\star)^4 \right ] \, ,
\eeq
for pure SU(3) and with $q^\star$ as above.
We can also obtain  $g^2_\MSb$ to third order from $g^2_P$, for which
we will use the BLM~\cite{BLM} improved relation 
\beq\label{gMSbofP}
  g^2_\MSb(\e^{-5/6} q^\star) \l= g^2_P(q^\star)
 \, \left [ 1 \+ {1\o 2\pi^2} ~ g^2_P(q^\star)^2
              \+  {0.9546933\o (4\pi)^2} ~ g^2_P(q^\star)^4 \right ] \, .
\eeq

In all four schemes the three-loop running of
the coupling is known (see~\cite{LW3loop,Alles} for the bare, 
e.g.~\cite{Alles} for the $E$-, \cite{V3loop} for the $V$-, 
and~\cite{MSb3loop} for the $\MSb$-scheme).
The scheme-dependent three-loop correction to the two-loop 
scaling function amounts to a multiplicative factor of the form
$1+c_S \, g_S^2$. The value of the coefficient $c_S$ in various schemes
is given in table~\ref{tab:lambda}, together with the $\La$-parameters.
Our ansatz for a three-loop fit of the string tension is therefore 
of the form
\beq\label{ansatz_3loop}
(a\sqrt{\si})(g_S(q)) \l=  \, (1 \+ c_S \, g^2_S(q))\, f(g^2_S(q)) \,
 \left[ 1 \+ c_2           \, \hat{a}(g_S(q))^2
          \+ c_4           \, \hat{a}(g_S(q))^4 \right ]
         /{\La_S\o a q \sqrt{\si}} \, .
\eeq
Ultimately, once better data for $r_0$ are available on fine lattices,
these fits should also be performed for $r_0$. At present, however, the
data for the string tension are significantly better, overall.

\begin{table}[tb] \centering
\begin{tabular}{ | c | r | c | }
\hline
Scheme & $c_S$~~~~~~ & $\La_\MSb/\La_S$ \\[0.7mm]\hline
 bare      & $0.189604$                   & 28.80934 \\
 $E$       & $0.011629$                   & 13.88031 \\

 $V$       & $-0.136923$                   & 0.625192 \\
 $\MSb$    & $-0.012631$                   & 1 \\
\hline
\end{tabular}      
\vskip 2mm
\caption{Three-loop coefficient $c_S = (b_1^2-b_0 b_{2,S})/(2b_0^3)$
         of the scaling function
         and $\La$-parameter for pure SU(3) gauge theory
         in various schemes.
}
\label{tab:lambda}
\vskip 7mm
\end{table}

\begin{table}[tb] \centering
\begin{tabular}{ | c | c | c | c | l | l | c | }
\hline
Scheme    & Order & $g^2_S(q)$ & $aq$ & ~~~$c_2$ 
                         & $\La_S/(aq\sqrt{\si})$ & $\La_\MSb$~(MeV)\\[0.7mm] 
\hline
 bare\rule{0mm}{4mm}
   & 2 & $6/\beta$       &  1     & $0.171(17)$&$0.01325(12)$& 178(2)\\[0.3mm]
   & 3 & $6/\beta$       &  1     & $0.157(16)$&$0.01550(14)$& 208(2)\\
\hline
 $E$\rule{0mm}{4mm}
   & 2 &$3(1-W_{11})$    &  1     & $0.080(6)$ &$0.03817(21)$& 246(1)\\[0.3mm]
   & 3 &$3(1-W_{11})$    &  1     & $0.078(6)$ &$0.03864(21)$& 249(1)\\
\hline
 $V$\rule{0mm}{4mm}
   & 2 & $g^2_P(q)$      & 3.4018 & $0.057(6)$ &$0.2604(13)$ & 258(1)\\[0.3mm]
   & 3 &eq.~\eqn{gVofP}  & 3.4018 & $0.130(8)$ &$0.2449(15)$ & 242(1)\\
\hline
$\MSb$\rule{0mm}{4mm}    
   & 2 &$g^2_P(q^\star)+{1\o 2\pi^2}g^4_P(q^\star)$
                         & 1.4784 & $0.074(6)$ &$0.3539(19)$ & 243(1)\\[0.3mm]
   & 3 &eq.~\eqn{gMSbofP}& 1.4784 & $0.072(6)$ &$0.3647(19)$ & 251(1)\\
\hline
\end{tabular}      
\vskip 2mm
\caption{Fit results for the string tension data $\beta\geq 5.6$ in
various schemes to an ansatz of the form~\eqn{ansatz_3loop} (the values
of the fit parameter $c_4$ are not shown). The order indicates which
scaling function was used (two- or three-loop) and, for the $V$- and
$\MSb$-schemes, to which order the coupling was extracted from the measured 
plaquette. We assume $\sqrt{\si}\s=465$~MeV in converting to physical
units. Details are in the text.
}
\label{tab:lamMSb}
\vskip 7mm
\end{table}

For each scheme we have performed fits using a two-loop and
a three-loop ansatz. For the $V$ and $\MSb$ cases the coupling is
then also extracted consistently to second, respectively, third
order from the measured value of the plaquette.
We find that all data can be fitted reasonably well in all schemes 
with either the two- or the three-loop ansatz. In either order the fits
in the different schemes are of comparable quality. 

When leaving out small $\beta$ points
the fit parameters are essentially stable within errors in all cases.
This is particularly impressive for $\La_S/( a q \sqrt{\si})$, which is
stable on the 1\% level even when all $\beta < 6.0$ points are left
out of the fit. By then $c_4$ has become ill-determined and can be
set to 0. But $c_2$ is still very significantly different from 0;
fixing it to 0 gives very small confidence levels (often astronomically
small; in the bare coupling scheme this is even true when leaving out the
$\beta\s=6.0$ points!). 

The stability of the fit parameters also holds when leaving out large
$\beta$ points. The $\La$-parameter is stable on the 1\% level even
when leaving out all $\beta\s=6.2, 6.4$ and $6.5$ points! This
clearly illustrates the quality of the fit ansatz.

We should mention that the reduced $\chi^2$
of the fits become very small ($\chi^2/N_{{\rm DF}} < 0.25$) once only
$\beta \geq 5.85$ points are included in the fits. We take this as an
indication that the errors on fine lattices have been, if anything, 
overestimated (at least for the data collected in table~\ref{tab:allsig}).

In table~\ref{tab:lamMSb} we show results for the various fits. For
definiteness we have chosen to fit $\beta=5.6 - 6.5$ in all
cases. We now see interesting differences between the schemes:
Most obviously, that the bare scheme has the largest $a^2$ coefficient
$c_2$, and shows a large increase in the $\La$-parameter when going
from two to three loops.\footnote{Note that the $a^2$ corrections
to the perturbative scaling of the string tension are not ``physical'', 
in the way that, say, $\Ord(a^2)$ violations of rotational symmetry
would be. 
In fact, we could perform a (non-perturbative) change of coupling to
the ``$\si$-scheme'', by defining $g^2_\si$ via
$(a\sqrt{\si})(g) \equiv f(g^2_\si)\, \sqrt{\si}/\La$. In this scheme
there would be no $\Ord(a^2)$ corrections whatsoever.}
The $E$- and $\MSb$-schemes, on the other hand, show much smaller 
$a^2$ corrections, and the $\La$-parameter is much more stable when going
from two to three loops. We regard the results in table~\ref{tab:lamMSb}
as a reconciliation of the ideas of Parisi-Lepage-Mackenzie with those
of Allton.

Note that the change in going from two to three loops is significantly
larger for the $V$- than for the $E$- or $\MSb$-schemes.
This can be traced back to the large value of the
three-loop coefficient in the scaling function, cf.~table~\ref{tab:lambda}.
This is also the reason for the jump in $c_2$.
What is a bit surprising, perhaps,
is that the results labeled $\MSb$, which really
come from a mix of the $\MSb$- and the $V$-scheme (the latter was
used to extract a physical coupling from the measured plaquette), show
much better agreement between the two and three loops results
than the $V$-scheme. 
In any case, one thing does seem clear, namely, that the values 
of $\La_\MSb$ extracted from the bare coupling scheme should not be 
trusted.

As another check of systematic errors we have considered using the
$1\!\times \!2$ Wilson loop, $W_{12}$, instead of $W_{11}$, to extract
$g^2_P(q^\star)$. Using the second order results of~\cite{HK} 
extrapolated to infinite volume and $q^\star = 3.0668/a$ in a two-loop
fit, we find that $\La_\MSb$ goes up by 16~MeV and 14~MeV, respectively,
compared to the results of table~\ref{tab:lamMSb} for the $V$- and
$\MSb$ schemes. 
One can also define an ``$E_{12}$''-scheme, using the expansion 
of $1-W_{12}$ analogously to eq.~\eqn{W11expn}. From~\cite{HK} we
%
obtain $\La_\MSb/\La_{E_{12}} \s= 27.4576$ for this scheme. 
Two-loop fits now show a decrease of $\La_\MSb$ by 15~MeV compared 
to table~\ref{tab:lamMSb}. 
Unfortunately we can not perform three-loop checks for the ``$12$''
schemes.

Our final check   of systematic errors consists of the
following worst case scenario. Let us assume that for some unknown
reason the true $\sqrt{\si}$ at both $\beta\s=6.4$
and $6.5$ are higher or lower by 2\% than the values in 
table~\ref{tab:allsig}. First of all we find that the $\chi^2$ of
our fits jump up by a factor of 2 or 3 (depending on the fitting 
interval), making the scenario appear unlikely indeed. 
Nevertheless, the value of $\La_\MSb$ changes only by about $1-2\%$.
This was to be expected, since, as mentioned earlier, we can leave
out the $\beta\s= 6.4, 6.5$ (and even $6.2$) points completely from the fit,
without significantly affecting the fit parameters.

As our central value for $\La_\MSb$ we will quote the average of the 
three-loop results in the improved coupling schemes. 
Among these schemes, the systematic errors appear to be largest in the
$V$-scheme; the difference between the two- and three-loop results, as well
as switching from $W_{11}$ to $W_{12}$ to extract the coupling, 
give a change in $\La_\MSb$ of 16~MeV. We will quote this
as our error:
\beq\label{LamMSb}
  \La_\MSb \l= 247(16)~{\rm MeV} \, .
\eeq
This is consistent with the value recently obtained by the ALPHA 
collaboration~\cite{LamALPHA}, as well as earlier results obtained
from the string tension (see e.g.~\cite{BS}).

To get~\eqn{LamMSb} 
we used a value of $\sqrt{\si} = 465$~MeV for the string tension.
This value is somewhat higher than the ones commonly
used in the past. However, several recent studies quote a value of
this size~\cite{Bliss,BSW}, and we would also like to make the following
simple, numerical observation.  The results in table~\ref{tab:allsig}
show that $r_0\sqrt{\si}$ is consistent, within errors of about 1\%, 
with the string picture value
\beq 
 r_0\, \sqrt{\si} ~=~ \sqrt{1.65-{\pi\o12} } ~=~ 1.178219 \, .
\eeq
 It is generally
believed~\cite{Sommer} that $r_0$ is not larger than about 0.5~fm.
Using $\sqrt{\si} = 1.178/r_0$ therefore gives 465~MeV (as a lower bound)
for $\sqrt{\si}$.

\section{Conclusions}

We have presented a high statistics study of the static potential in
pure lattice gauge theory, using the Wilson plaquette action. 
These data and improved analysis strategies have enabled us to give
very precise determinations of the string tensions and Sommer \hbox{(-type)}
scales for $\beta \leq 6.0$, i.e.~on lattices of spacings of about
0.1~fm and above.
Combining our results with previous determinations,  $a\sqrt{\si}$
is now known to   1\% 
in the range 
$5.54 \leq \beta \leq 6.5$. According to our and other studies
(quoted in sect.~4) this error also includes possible finite
volume effects.

The scale $r_0$ is presently known somewhat less accurately on the
finest lattices. Using our method to determine Sommer-type scales it
should be straightforward to determine $r_0$ for all lattice spacings
at least as accurately as the string tension. In particular, any
remaining finite volume effects should be much smaller for $r_0$.
Needless to say,
our method can also be used for other kinds of gauge actions.

We have employed the ansatz of Allton~\cite{Allton} to provide  
parameterizations of $\sqrt{\si}$ and $r_0$ that are accurate to
about 1\% 
for $5.6 \leq \beta \leq 6.5$.  By considering fits of
this kind for the string tension in various coupling schemes we have 
estimated the $\La$-parameter of quenched QCD to be
$\La_\MSb \s= 247(16)~{\rm MeV}$. We concluded that when it comes
to extracting physical parameters, instead of just fitting data, the
improved coupling schemes are very much superior to the bare coupling
scheme, even when corrected for $a^2$ effects as suggested by Allton.

The results presented here should be useful in high precision
scaling checks, comparisons of different actions, etc. Our motivation 
for this study has, in fact,  been the desire to provide a
significant check of the Symanzik improvement program by performing
a scaling check of the rho mass in units of the string tension
for the SW action with a non-perturbatively determined clover 
coefficient~\cite{EHK}.  This is not possible without accurate
knowledge of the string tension (or $r_0$).
Previous estimates of the string tension would have lead
to misleading conclusions about the scaling behavior of the rho mass
or other observables.

\vskip 13mm

\noindent
{\bf Acknowledgements}

\noindent    
This work is supported by DOE grants DE-FG05-85ER25000 and 
DE-FG05-96ER40979.
The computations in this work were performed on the workstation
cluster, the CM-2, and the new QCDSP supercomputer at SCRI. 

\newpage


\end{document}